\documentclass[fleqn,usenatbib]{mnras}

\usepackage{newtxtext,newtxmath}

\usepackage[T1]{fontenc}

\DeclareRobustCommand{\VAN}[3]{#2}
\let\VANthebibliography\thebibliography
\def\thebibliography{\DeclareRobustCommand{\VAN}[3]{##3}\VANthebibliography}

\usepackage{graphicx}	
\usepackage{amsmath}	

\usepackage{isotope}
\usepackage{siunitx}
\usepackage{wrapfig,framed}
\usepackage{outlines}

\newcommand{\SLR}
{short-lived radioisotope}

\newcommand{\SLRs}
{short-lived radioisotopes}

\newcommand{\Min}
{ _\textit{min} }

\newcommand{\Max}
{ _\textit{max} }

\newcommand{\acc}
{ _\textit{acc} }

\newcommand{\mean}
{ _\textit{mean} }

\newcommand{\unenr}
{_\textit{ne}}

\newcommand{\crit}
{ \text{crit} }

\usepackage[dvipsnames]{xcolor}
\usepackage{ulem}

\title[ ]{Prevalence of short-lived radioactive isotopes across exoplanetary systems inferred from polluted white dwarfs}

\author[]{Alfred Curry$^{1,2}$, Amy Bonsor$^{2}$, Tim Lichtenberg$^{3}$, Oliver Shorttle$^{2,4}$
\\
$^{1}$ Astrophysics Group, Imperial College London, Blackett Laboratory, Prince Consort Road, London SW7 2AZ, UK\\
$^{2}$ Institute of Astronomy, University of Cambridge, Madingley Road, Cambridge, CB3 0HA, UK\\
$^{3}$ Atmospheric, Oceanic and Planetary Physics, Department of Physics, University of Oxford, Parks Road, Oxford, OX1 3PU, UK\\
$^{4}$ Department of Earth Sciences, University of Cambridge, Downing Street, Cambridge, CB2 3EQ, UK
}

\date{Accepted  XXX. Received YYY; in original form ZZZ}

\pubyear{2022}

\begin{document}
\label{firstpage}
\pagerange{\pageref{firstpage}--\pageref{lastpage}}
\maketitle

\begin{abstract}
In the Solar System short-lived radioisotopes, such as $^{26}\text{Al}$, played a crucial role during the formation planetary bodies by providing a significant additional source of heat. Notably, this led to early and large-scale melting and iron core formation in planetesimals and their loss of volatile elements, such as hydrogen and carbon. In the context of exoplanetary systems, therefore, the prevalence of short-lived radioisotopes is key to interpreting the observed bulk volatile budget and atmospheric diversity among low-mass exoplanets. White dwarfs that have accreted planetary material provide a unique means to infer the frequency of iron core formation in extrasolar planetesimals, and hence the ubiquity of planetary systems forming with high short-lived radioisotope abundances. Here, we devise a quantitative method to infer the fraction of planetary systems enriched with short-lived radionuclides upon planetesimal formation from white dwarf data. We argue that the current evidence from white dwarfs point towards a significant fraction of exo-planetesimals having formed an iron core. Although the data may be explained by the accretion of exo-moon or Pluto-sized bodies that were able to differentiate due to gravitational potential energy release, our results suggest that the most likely explanation for the prevalence of differentiated material among polluted white dwarfs is that the Solar System is not unusual in being enriched in $^{26}$Al. The models presented here suggest a ubiquitous pathway for the enrichment of exoplanetary systems by short-lived radioisotopes, disfavouring short-lived radioisotope enrichment scenarios relying on statistically rare chance encounters with single nearby supernovae, Wolf-Rayet or AGB stars.
\end{abstract}




\section{Introduction}
Short-lived radioactive isotopes have dual importance in planetary science.  In the Solar System, \SLRs\: are powerful chronometers dating the timing of physical and chemical processes during planetary formation \citep{2014prpl.conf..571G}.  {In particular, \isotope[26]{Al}, which has a half-life of $\SI{7.17e5}{}$ yrs \citep{ALhalf-norris83,Alhalf-wu}, } is widely used to date various extraterrestrial materials relative to the formation of Calcium, Aluminium-rich inclusions (CAIs), the earliest dated solids that formed during the formation of the proto-Sun \citep{2014prpl.conf..809D}. However, \SLRs\: are not only passive geochronometers, they play an active role in planet formation processes through the heat they release on decay. \isotope[26]{Al} is a powerful heat source for planetary objects that are formed within a few half-lives after CAIs, which explains early core-mantle differentiation in the small parent bodies of meteorites (planetesimals) $\sim$0.1--0.3 Myr after CAIs \citep{2006E&PSL.241..530S}. This internal heating and resulting geophysical and geochemical evolution may also be responsible for the chronological split in core ages and aqueous alteration between inner and outer Solar System planetesimals \citep{2021Sci...371..365L}, degassing of highly and moderately volatile species \citep{gilmour2009_icarus,2019GeCoA.260..204S,2019Icar..328..287W,2021PNAS..11826779H}, and vaporization of rock-forming elements \citep{2019Icar..323....1Y,2020Icar..34713772B} from accreting planetesimals. The key question we address in this manuscript is whether such \SLR-driven thermal processing has also been common in exoplanetary systems.

Within the Solar System, the presence of short-lived radioactive isotopes, particularly \isotope[26]{Al}, led to the extensive thermal processing of even relatively small asteroids \citep[$\gtrsim 10$ km,][]{2006M&PS...41...95H} that formed within the first few Myrs after CAIs.  Melting of small bodies can only occur with heat from \SLRs: whilst the release of gravitational potential energy during the formation of large planetary bodies ($\gtrsim$ 1,500 km) can drive large-scale melting on its own \citep{Elkins-Tanton12-review}, the heating of small asteroids by potential energy release is insignificant.  The ability for small planetesimals to melt early during the planet formation era is particularly important for the eventual composition of Earth to super-Earth mass planets, assuming they grow by planetesimal accretion.  Small bodies do not have strong enough gravity to retain outgassed volatiles, so their melting drives volatile species such as water and carbon dioxide out of the building blocks of planets.  Abundant $^{26}$Al during rocky planet formation may therefore alter the chemical bulk abundances of volatile and atmophile elements across extrasolar planetary systems by orders of magnitude \citep{2021ApJ...913L..20L,2021ApJ...919...10A}, and therefore influence their long-term climate and surface conditions \citep{2016SSRv..205..153M,2018A&ARv..26....2L}. Exoplanet observations in the 2020s will have the potential to probe the atmospheres and climates of select exoplanets on short-period orbits. To understand the history of these planets' atmospheres will require independent constraints on the importance of \SLR\: heating during planet formation throughout the galaxy \citep{2021arXiv211204663W,2022arXiv220310023L}.

While the elevated abundance of \isotope[26]{Al} in the forming Solar System is established, its precise origin is debated.  Two observations pose a challenge to understanding the Solar System's initial inventory of \isotope[26]{Al}: its half-life ($\SI{7.17e5}{}$ yrs) is short relative to typical mixing timescales between the interstellar medium (ISM) and giant molecular clouds GMCs, from which stars form \citep[$\sim 100$ Myr]{deAvillez2002}; and, the initial Solar System enrichment is $\sim$3--25 times that of the galactic mean \citep{lugaro18-review}.  Together these observations rule out the unmodified interstellar medium as a source of the Solar System's \isotope[26]{Al} \citep[][]{Meyer2000,Benoit19}. Instead, a more local origin of \isotope[26]{Al} is required \citep{2016ARA&A..54...53N}. \isotope[26]{Al} can be produced in both supernovae and the winds of massive stars \citep[][]{lugaro18-review}, and models of delivery either rely upon a close encounter with such an event or sufficient mixing in the star forming region \citep{2020RSOS....701271P}. The former implies that the Solar System is relatively rare, $\lesssim 1\%$, while the later implies that it is more common $\sim 25\%$. In the former case the chemical abundances in atmophile and moderately volatile elements in the terrestrial planets would be interpreted a chance event and highly volatile-rich exoplanets may be the norm across the galaxy. In the latter case, the distribution between volatile-rich `ocean' worlds and drier, terrestrial-like planets may be statistically traceable across the exoplanet population with near-future transit surveys \citep{2019NatAs...3..307L}.

The melting history of planetary building blocks can also be tested more directly from white dwarf observations.  White dwarfs that have accreted planetary material provide a unique means to assess whether or not small planetary bodies underwent the large-scale melting necessary to form an iron core. White dwarfs are the faint remnants of most planet-hosting stars and provide an ideal laboratory for studying the geology of exoplanetary bodies. Accreted planetary material shows up in the atmospheres of 30-50\% of white dwarfs \citep{Koester2014, ZK10}. The otherwise clean spectra show metallic features from the accreted material, that sink out of sight on timescales of days (hot, 20,000 K DAs) to millions of years (cool, 6,000 K DZs). High (low) Fe abundances suggest that some white dwarfs have accreted a core-rich (mantle-rich) fragment of a larger planetary body \citep[e.g.,]{Melis2011, Gaensicke2012, Wilson2015}.  

The white dwarf observations have the unique ability to probe the prevalence of core--mantle differentiation across exoplanetary systems \citep{Bonsor2020}. In the Solar System, \SLRs\: are a key source of energy to fuel core--mantle differentiation. It is, however, not clear how common it is for exoplanetary systems to have a significant budget of \SLRs. In this work we describe a method to link \SLR\: enrichment to the level of core--mantle differentiation in planetary systems, as measured through white dwarfs. Thus, the observations of core--mantle differentiation observed in white dwarfs can be used to probe the frequency of \SLR\:  enrichment across the galaxy, as suggested by \cite{JuraYoung}. 

{ Since the white dwarfs simply probe planetesimal heating, they are agnostic as to the particular \SLR\: causing differentiation. In the solar system, \isotope[26]{Al} was the most significant, hence our focus on it and its extensive study in the literature. In principle, however, our work also applies to \isotope[60]{Fe} which can act as a major heat source if present in sufficient quantities.}

\S\ref{sec: Method} describes our analytic framework and key assumptions, and \S\ref{sec: linking_evidence} applies the model to current observations to deduce the prevalence of \SLRs\: across exoplanetary systems. \S\ref{sec: discuss} discusses how robust our conclusions are and their implications for the field in light of observational uncertainties. \S\ref{sec: conclusion} summarises our conclusions.

\section{An analytic model for the frequency of (iron) core formation across exoplanetary systems} \label{sec: Method}

\subsection{Model outline}

\begin{figure}
	\centering
	\includegraphics[width = \linewidth]{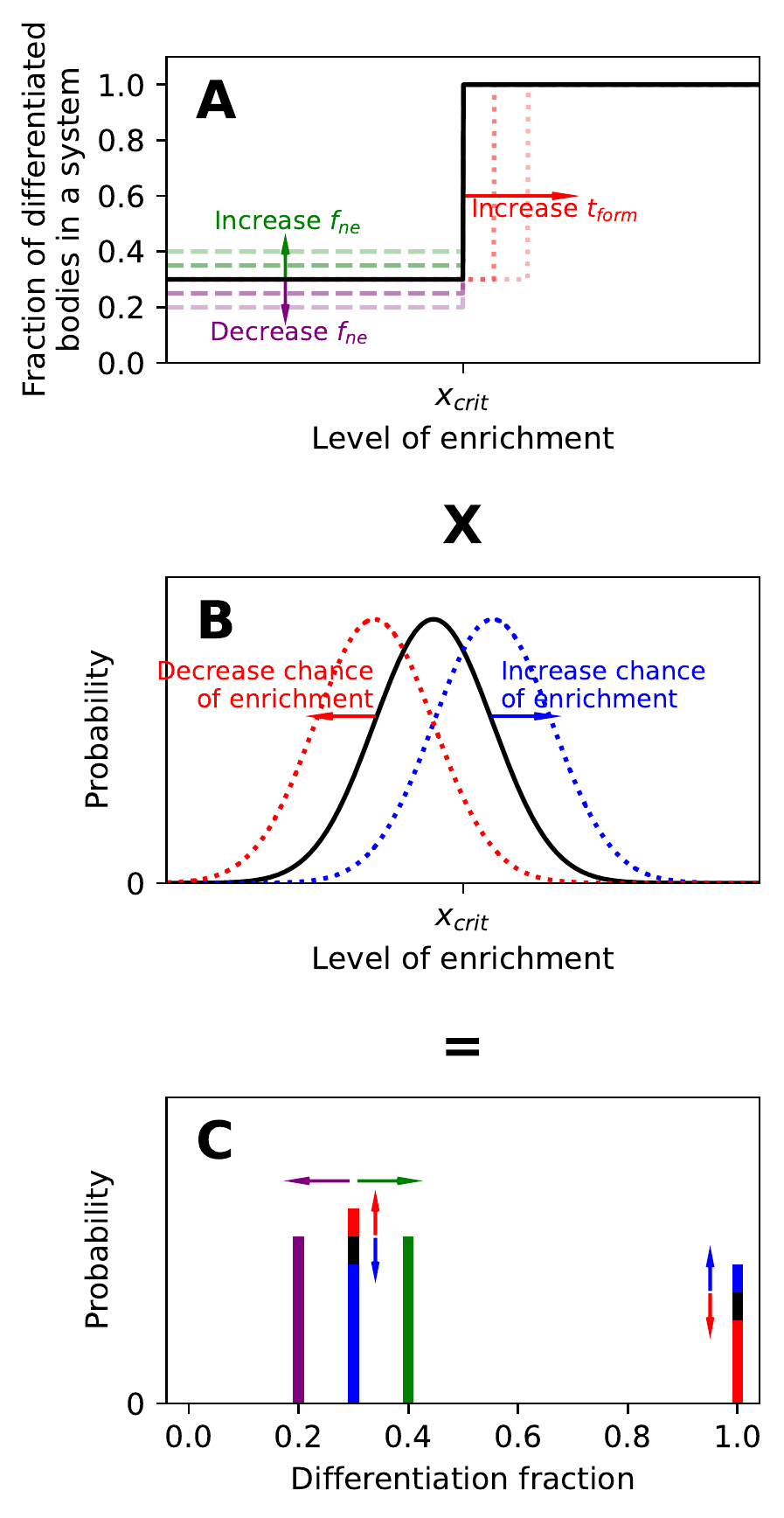}
	\caption{{ Schematic showing the steps in our model for linking the observed fraction of bodies that are differentiated in systems to the probability of a system being enriched with \SLRs. This is a graphical representation of Eq. \ref{eq: genfmean}. Panel A shows the fraction of planetary bodies in a system that are differentiated as a function of how enriched the system is in short-lived radioisotopes. In our model all planetary bodies are enriched in systems with $x>x_{\rm crit}$ (\S\ref{sec: R_SLR}), although this is a function of formation time. To find the distribution across all systems (Panel C) this must be multiplied by the enrichment probability, which is shown in Panel B. For illustrative purposes this is a normal distribution, but need not be in reality, and will depend on the dominant enrichment mechanism. The result of this model is that two populations are produced in Panel C, a with high and low differentiation fraction, for enriched and non-enriched systems. Colours in Panel C correspond to the labels in the earlier panels, showing how results change for changing parameters. An increase in formation time, $t_\text{form}$, has the same effect as lowering the chance of enrichment so both labels are the same colour. In order to find the observable mean fraction of differentiated bodies, the mean of the populations is taken (Eq. \ref{eq: simplefmean})}}\label{fig: fxform}
\end{figure}

The aim of this paper is to construct a model that links the probability of enrichment by \SLRs\: to the fraction of planetary bodies, for example exo-asteroids in a planetesimal belt, that form an iron core.  Observations of core- or mantle-rich material in the atmospheres of white dwarfs shed light on the fraction of planetary bodies that form an iron core. However, the white dwarfs sample this fraction averaged across all exoplanetary systems, which we denote $f\mean$. 

{We show an outline of our model schematically in Fig. \ref{fig: fxform}, and it is explained in more detail in the following sections. 

We start by considering the fraction of bodies in an individual system that form an iron core.
This function, $f (x, n, t)$, }depends upon, (i) the level of \SLR\: enrichment in that planetary system, $x$, which provides a heating source for large-scale melting and core formation, (ii) the size distribution of the planetary bodies, $n(R)$, as the largest may form iron cores without the need for \SLRs\: (\S\ref{sec: R_SLR}), and (iii) the distribution of times for planetesimal formation, $t$, after \SLR\: enrichment of the system. 

In each individual planetary system both early-formed, and thus sufficiently $^{26}$Al-enriched, planetary bodies and the largest planetary bodies, where gravitational potential energy alone provides sufficient heating to lead to large-scale melting, can contribute to iron core formation. Planetary bodies that form too late and are too small will not form iron cores. {For simplicity, we here consider a single epoch of (averaged) planetesimal formation time per system, but note that this may not reflect how planetesimal formation occurs in nature.

The function we use is a step function as shown in Panel A of Fig. \ref{fig: fxform}, where in systems above a critical \SLR\: enrichment level, $x_\crit$, all bodies are differentiated, and in those below $x_\crit$ only the largest bodies are. We justify this further in \S\ref{sec: R_SLR}.

In order to find the mean fraction of bodies that have differentiated also requires knowing the proportion of systems that are enriched in \SLRs. We consider this a function, $P(x, n , t)$: the probability that a system has enrichment $x$, size distribution $n$ and planetesimals forming with average formation time $t$, although we will show in \S\ref{sec: simplified} that the exact form is not important. An example function is shown in Panel B of Fig. \ref{fig: fxform}.

Therefore,} the mean fraction of planetary bodies that are differentiated, $f\mean$, is simply given by the weighted average over all possible enrichment levels, formation times and size distributions,
\begin{equation}
	f\mean = \int_{x}\int_{n}\int_{t} P(x, n , t) \: f(x,n,t) \: \mathrm{d}x \:\mathrm{d}n \:\mathrm{d}t  \label{eq: verygenfmean},
\end{equation}
{We note here that the average loses some information. Our model predicts two populations, as demonstrated in Panel C, one with high fractions of differentiation, due to having been enriched with SLRs\: and a lower population that was not enriched, but the mean necessarily averages over this.} 

For the purposes of this work, we assume that the formation time and size distribution in a system is independent of the enrichment level in a given system and does not vary between systems. The limits and shortcomings of these assumptions are discussed in \S\ref{sec: discuss}. As a result of these assumptions, Eq. \ref{eq: verygenfmean} simplifies to
\begin{equation}
	f\mean = \int^{\infty}_{x=0} P(x) \: f(x,n,t) \: \mathrm{d}x  \label{eq: genfmean},
\end{equation}
where $n$ and $t$ are now a representative size distributions and time of formation and $P(x)$ is the probability that an individual planetary systems is enriched up to a level $x$. In the following section we discuss the form of $f(x,n,t)$. A guide to our notation can be found in Table \ref{tab: params}.

\begin{table}
\resizebox{\columnwidth}{!}{%
\begin{tabular}{l|p{0.7\columnwidth}}
Parameter             & Definition   \\ \hline
$x$                   & Short-lived radioisotope enrichment level in a system, at some early time following enrichment, $t_0$.  \\
$t$                   & Average formation time, measured relative to $t_0$. \\
$R$                   & Planet/planetesimal radius.                                     
\\
$n(R)$                & Size distribution in a given planetary system -  the number of bodies with radius between $R$ and $R + \mathrm{d}R$.  \\
$P(x)$                & Probability of a system having \SLR\: enrichment between $x$ and $x + \mathrm{d}x$. \\
$f(x,n,t)$            & Fraction of bodies that are have iron cores in a system with \SLR\:enrichment $x$, size distribution $n$ and average formation time $t$. \\
$f\mean$              & Mean fraction of bodies that are form iron cores across all systems that can pollute white dwarfs. Equivalently, the fraction of the total number of bodies that can pollute white dwarfs that are core-mantle differentiated. 
\\
$R\acc$               & Radius above which bodies form an iron core due to energy released during accretion.
\\
$R_\text{SLR}(x,t)$ & Radius above which bodies form an iron core due to energy released by \SLRs.                                       \\
$\tau_\text{SLR}$   & Mean lifetime of the relevant \SLR, generally \isotope[26]{Al}{, which has $\tau_\text{26Al} = 1.03$ Myr.}                   \\
$x_\crit$              & Critical \SLR\: enrichment level, above which core formation occurs.                                         \\
$x_\crit(t)$           & The initial \SLR\: enrichment level, at $t = t_0$, above which planets will form iron cores if formed at time $t$. $x_\crit(t) = x_\crit e^{t/\tau_\text{SLR}}$.  \\
$P_e(t)$            & Probability of having \SLR\: enrichment above the critical level at formation time $t$.                             \\
$f\unenr$             & Fraction of bodies that are core-mantle differentiated in systems with \SLR\: enrichment lower than the critical value \\
$q$                   & Size distribution index, $n(R) \propto R^{-q}$.                          \\
$R_\textit{min/max}$       & The minimum/maximum formation radii of bodies that go on to pollute white dwarfs.
\end{tabular}%
}
\caption{A list of definitions of parameters used in our model.}\label{tab: params}
\end{table}

\subsection{Fraction of differentiated planetesimals} \label{sec: diff_frac_sys}

If planetary bodies are to form an iron core, they must undergo a phase of large-scale melting. This melting leads to the differentiation of the planetary body. The heating can be powered by two potential sources: (i) Potential energy released during accretion. (ii)  Energy released during radioactive decay of short-lived radioisotopes. Both of these mechanisms depend on the {radius that the bodies grow to during accretion} and hence the number of bodies in a system that are differentiated will depend on the {initial size distribution of bodies in that system.}

The first heating mechanism in our model depends solely upon the size, with only the largest planetary bodies having sufficient gravitational potential energy to melt. Therefore, this can be represented by a cut-off size, $R\acc$, where all bodies larger than $R\acc$ are core--mantle differentiated. This cut-off radius is of order 1000 km, as discussed in the next section. 

The second threshold depends on both the amount of radionuclide heating and the size of the planetary bodies, with the smallest bodies containing insufficient \isotope[26]{Al} to fuel core--mantle differentiation. This time the cut-off size, $R_\text{SLR}(x,t)$, is much smaller ($\sim10$s km, see \S\ref{sec: R_SLR} and Fig. \ref{fig: Licht_fracs}) and depends critically on the initial level of enrichment, and time of formation. These two mechanisms for differentiation are explained below in greater detail.

\subsubsection{Minimum radius for differentiation due to release of gravitational potential energy} \label{sec: R_acc}
When planets form the material must lose gravitational energy and this must be released and either heat up the planet or be radiated away. The minimum size for which there is sufficient accretional energy to lead to large-scale melting can be estimated by considering the available gravitational potential energy. Following \cite{Elkins-Tanton12-review} (\S3.2), assuming a constant density, this energy is 
\begin{equation}
    E_g = - \frac{4\pi}{5} M \rho G R^2, 
    \label{eq:eg}
\end{equation}
where $M$ is the mass of the planetesimals, $\rho$ its density, and $R$ its radius. The temperature of the body increases as 
\begin{equation}
    \Delta T = \frac{E_g}{M C_P}, 
\end{equation}
where $C_P$ is the specific heat capacity.  Combined with Eq.~\ref{eq:eg}, the minimum size at which a temperature rise of $\Delta T_\textit{\rm melt}$ occurs is given by: 
\begin{equation}
    R\acc = \sqrt{ \frac{5}{4\pi} \frac{ C_P \Delta T_\textit{\rm melt}}{\rho G}}.
\end{equation}
Using conservative values of $\rho = \SI{4000}{\kg\:\m^{-3}}, C_P = \SI{1000}{\J \: \kg^{-1}\: \K^{-1}}$ \citep{SOLOMATOV-chapter}, and a temperature to allow sufficient melting of $\sim \SI{1600}{\K}$ \citep[][peridotite liquidus]{Lesh+Sper15} gives a radius of $R_{\rm acc} \sim \SI{1500}{\km}$. Thus, all planetary bodies larger than $\sim$1500km are assumed to form an iron core.

As a caveat, we note here that this picture is potentially over simplistic. In reality this energy is deposited as planets grow, which is a stochastic process \citep{Elkins-Tanton12-review}. It is also non-homogeneous, as giant impacts and secondary accretion will melt from the surface inwards. {To accurately model this process one needs to consider the combination of thermal evolution and accretion timescales \citep[e.g.,][]{Sramek12,sturtz2022structure}, since the planet can cool between impacts or growth from pebbles. On the other hand we neglect the extra energy released by core formation itself, where denser material sinks deeper into the potential well, meaning once a certain amount of melting is generated it may be easier to entirely differentiate \citep{Monteux09}. Overall, however, we consider our value as a lower estimate, due to the cooling of planets.}

\subsubsection{Minimum radius for differentiation due to \SLRs} \label{sec: R_SLR}
\begin{figure}
    \centering
    \includegraphics[width=\linewidth]{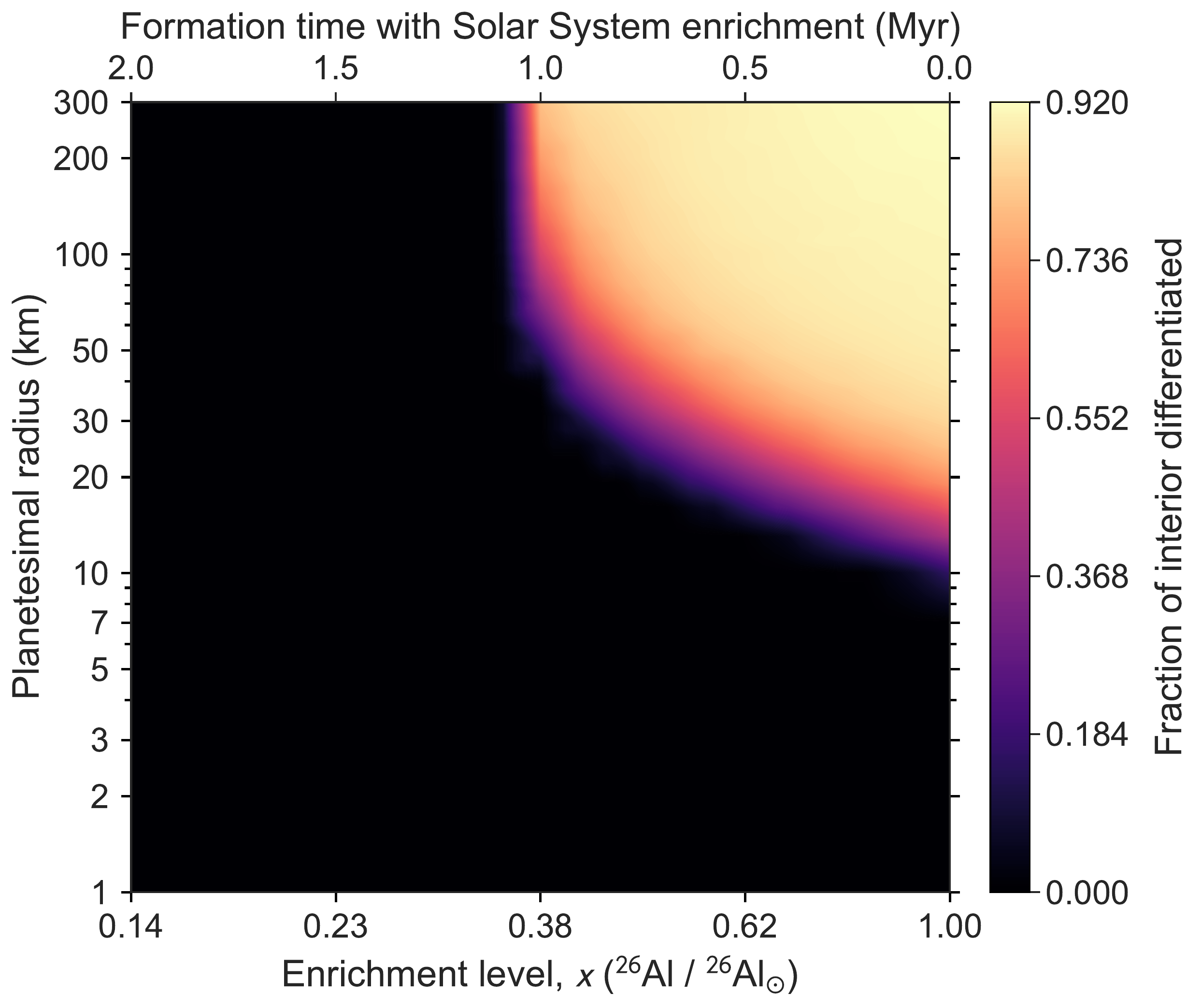}
    \caption{The fraction of planetesimals that melt sufficiently for metal (Fe) and silicates to segregate from one another and form an iron core, as a function of the planetesimal size and level of \isotope[26]{Al} enrichment $x$ \citep{2021Sci...371..365L}. All bodies in the top right-hand corner ($\gtrsim$ 10 km in radius and and $\gtrsim$ 0.38 $^{26}$Al/$^{26}$Al$_{\odot}$) will form an iron core. Enrichment is measured relative to the Solar System value at the formation of CAIs (Ca,Al-rich inclusions), the earliest known solids: (\isotope[26]{Al}/\isotope[27]{Al})$_\text{CAI}$ $= \SI{5.25e-5}{}$ \citep{26Al_Kita}. The top axis shows the time after CAI formation which corresponds to a given enrichment level.}
    \label{fig: Licht_fracs}
\end{figure}

For a given level of enrichment in \SLRs, planetary bodies below a certain size will never undergo large-scale silicate melting, as there is not enough energy, even from the decay of all their \SLRs\:  to heat the body to a sufficient temperature to melt. Depending on the level of enrichment in a given planetary system, as well as the time at which a given planetesimal forms, the minimum size for large-scale melting and iron core formation varies. Considering a system-averaged epoch of planetesimal formation, the simplest possible model would be to consider that all planetesimals in a planetary system enriched above a critical level, $x_{\rm crit}$, form an iron core, whilst no planetesimals in a system enriched below this level form an iron core. This can be justified, firstly, as in order to be detected, white dwarfs tend to accrete planetary bodies significantly larger than the limit and secondly, as will be shown, the transition occurs over a small range of planetesimal radii.  

In order to find the minimum radius for differentiation in the presence of \SLRs, $R_\text{SLR}(x,t)$, we follow the method of \cite{2016Icar..274..350L,2021Sci...371..365L}. In these models a single planetesimal is simulated using a 2-D fluid dynamical model \citep{2007PEPI..163...83G} with an initial enrichment level $x$ of \isotope[26]{Al} and a fixed radius. The composition of the planetesimals is assumed to be chondritic, melting and deformation of the rock are self-consistently evolved. In Fig. \ref{fig: Licht_fracs} the simulation results from 700 single-planetesimals simulations are linearly interpolated on the $x$-$R$ grid. To derive the fraction of each planetesimal body that undergoes metal-silicate differentiation during the decay of \isotope[26]{Al}, we conservatively assume that core formation only occurs for parts of the planetesimal that melt beyond the disaggregation treshold of silicate rocks ($>$ 40\%), when molten Fe or Fe-S droplets rain out from the internal magma ocean.

Fig. \ref{fig: Licht_fracs} shows that for \isotope[26]{Al} enrichment levels $\lesssim 0.38^{26}Al_\odot$ no planetesimals of any size will differentiate. Above that level, as one would expect, as the level of enrichment, $x$, increases the radius required to differentiate decreases. Fig. \ref{fig: Licht_fracs} shows that all bodies above 10km in radius will form an iron core at $^{26}Al_\odot$, whilst all bodies above 50 km will form an iron core at $0.38^{26}Al_\odot$. Thus, a good model for the behaviour seen here, is that all planetesimals in system enriched above, $x_{\rm crit} = 0.38^{26}Al_\odot$, form an iron core, if the sizes of bodies considered are larger than $\sim \SI{50}{\km}$. We return to this point in \S\ref{sec: valid_crit}.

{These models are run for solar system like conditions where \isotope[26]{Al} is the dominant radioactive heating source at these times. If in other systems other radioisotopes, such as \isotope[60]{Fe},  replace \isotope[26]{Al} as the dominant source of heating then $x_{\rm crit}$ simply represents the equivalent abundance. If a different \SLR\: was dominant it might affect timing slightly, due to half-life differences, but not the overall picture.}

This means that in exoplanetary systems with enrichment levels below $x_{\rm crit}$, only the largest planetary bodies form an iron core. In those systems that are enriched above the critical level ($x>x_{\rm crit}$), all planetary bodies that form sufficiently early are enriched. With a single averaged epoch of planet formation, this implies that either all planetary bodies form an iron core, or just the largest. This step function can be seen in the first panel of Fig. \ref{fig: fxform} and can be expressed as
\begin{equation}
    f(x) = 
    \begin{cases}
        f\unenr,& \text{if } x<x_\crit(t)\\
    1,              & x > x_\crit(t).
\end{cases} \label{eq: f_xform}
\end{equation}
Here we define
\begin{equation}
    x_\crit(t) = x_\crit e^{t/\tau_\text{SLR}} \label{eq: x_crit(t)},
\end{equation}
the amount of enrichment in the system initially so as to differentiate bodies formed at a later time $t$, when the heating produced by radioisotopes will have decreased due to decay. $f\unenr$ denotes the fraction of bodies that are differentiated in systems with {\it no enrichment} and is calculated by considering the size distribution.

\subsubsection{Size distribution} \label{sec: size_dist}
\begin{figure}
	\centering
	\includegraphics[width = \linewidth]{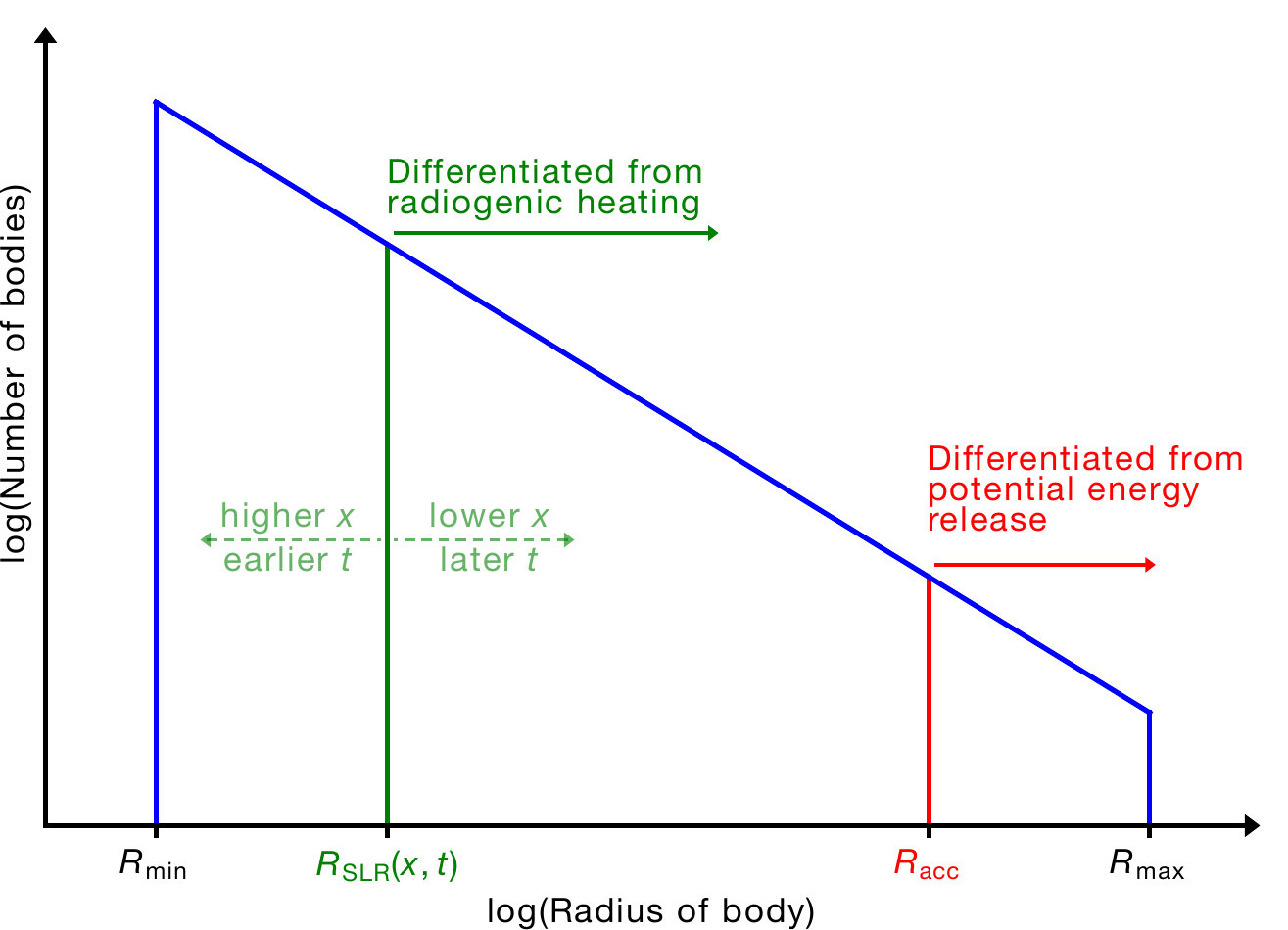}
	\caption{Schematic of the size distribution in a planetesimal belt with minimum planetesimal radii for core-mantle differentiation indicated by the two mechanisms. Planetary bodies larger than $R_\textit{acc}$ form an iron core due to accretional heating alone. Thus, in a planetary system that is not enriched, it is only planetesimals larger than $R_{\rm acc}$ that are core--mantle differentiated ($f\unenr$ in Eq. \ref{eq: simplefmean}). Planetary bodies larger than $R_\text{SLR}$ form an iron core due to heating from \SLRs.  $R_\text{SLR}$ increases if a system is less enriched ($x$) or planetesimals form later ($t$). If white dwarfs are polluted by planetesimals scattered inwards, the dynamical mechanisms at play are independent of size and will select at random from this size distribution.}\label{fig: cartoon}
\end{figure}
We assume the fraction of white dwarfs that accrete core--mantle differentiated planetary bodies (or fragments of them) to be simply the fraction of planetesimals in the belt that are core--mantle differentiated. This can be calculated by considering the size distribution, depicted in Fig. \ref{fig: cartoon}. Planetesimals larger than $R_{\rm acc}$ always form an iron core. If the planetary system is sufficiently enriched in \SLRs, planetesimals above $R_\text{SLR}$ also form an iron core. Thus, the number of planetesimals larger than these limits as a function of the size distribution is key to this model. This size distribution incorporates both the size distribution of planetary bodies scattered onto white dwarfs and the size distribution of planetary bodies when they initially formed. For simplicity, the size distribution is assumed to be a power law of the form
\begin{equation}
    n(R) \propto R^{-q},
\end{equation}
where the index $q$ is essentially a free parameter. If the system is in collisional equilibrium, $q\sim 3.5$. The size distribution produced by the streaming instability is slightly shallower \citep[$q=2.8$,][]{Simon16}. The present-day asteroid belt has $q \approx 4.5$ above 60 km in radius and hot Kuiper belt objects $q \approx 5$--$6$ \citep{2014ApJ...782..100F}. If runaway growth were to dominate the size distribution, with no subsequent collisional evolution, it is plausible that the size distribution could be as steep as $q\sim 6$ \citep[e.g.,][]{Makino98-runaway,Kokubo96-runaway}. For probing reasonable end-member scenarios of planetary growth in inner planetary systems, we proceed with values for $q$ of 3 and 6.

\subsection{The mean fraction of core--mantle differentiated planetesimals }\label{sec: simplified}

As all planetesimals in planetary systems enriched above a level, $x_{\rm crit}$, are assumed to form an iron (Eq. \ref{eq: f_xform}), the mean fraction of planetesimals that form an iron core across all exoplanetary systems, given by the integral in Eq. \ref{eq: genfmean}, can now be expressed as: 

\begin{equation}
	f\mean = \int^{x_\crit(t)}_{x=0} P(x) \:f\unenr \: \mathrm{d}x + \int^{\infty}_{x=x_\crit(t)} P(x) \times 1 \: \mathrm{d}x,  
\end{equation}
If we define
\begin{equation}
    P_e(t) =  \int^{\infty}_{x=x_\crit(t)} P(x) \mathrm{d}x \label{eq: P_enr(t)}
\end{equation}
then, using the fact that, by definition, $\int^{\infty}_{x=0} P(x) \mathrm{d}x =1$,
\begin{equation}
	f\mean = (1-P_e(t))f\unenr + P_e(t) \label{eq: simplefmean}.
\end{equation}
This very simple equation incorporates the important principles for this work. First, there are two contributing sources to the total number of differentiated bodies in systems: those due to large bodies with sufficient energy from accretion (first term) and those due to radioactive decay (second term). Second, the exact form of $P(x)$ is no longer crucial; it is now the probability of a system being enriched above the critical value, $P_e(t)$, that is important. Third, the time of planetesimal formation is crucial.  If planets (planetesimals) formed later, then the probability that there was sufficient \SLR\: enrichment at the birth of the system to drive differentiation is lower (Eqs. \ref{eq: P_enr(t)} and \ref{eq: x_crit(t)}), and thus, there will be fewer planetesimals with an iron core. 

\begin{figure}
	\centering
	\includegraphics[width = \linewidth]{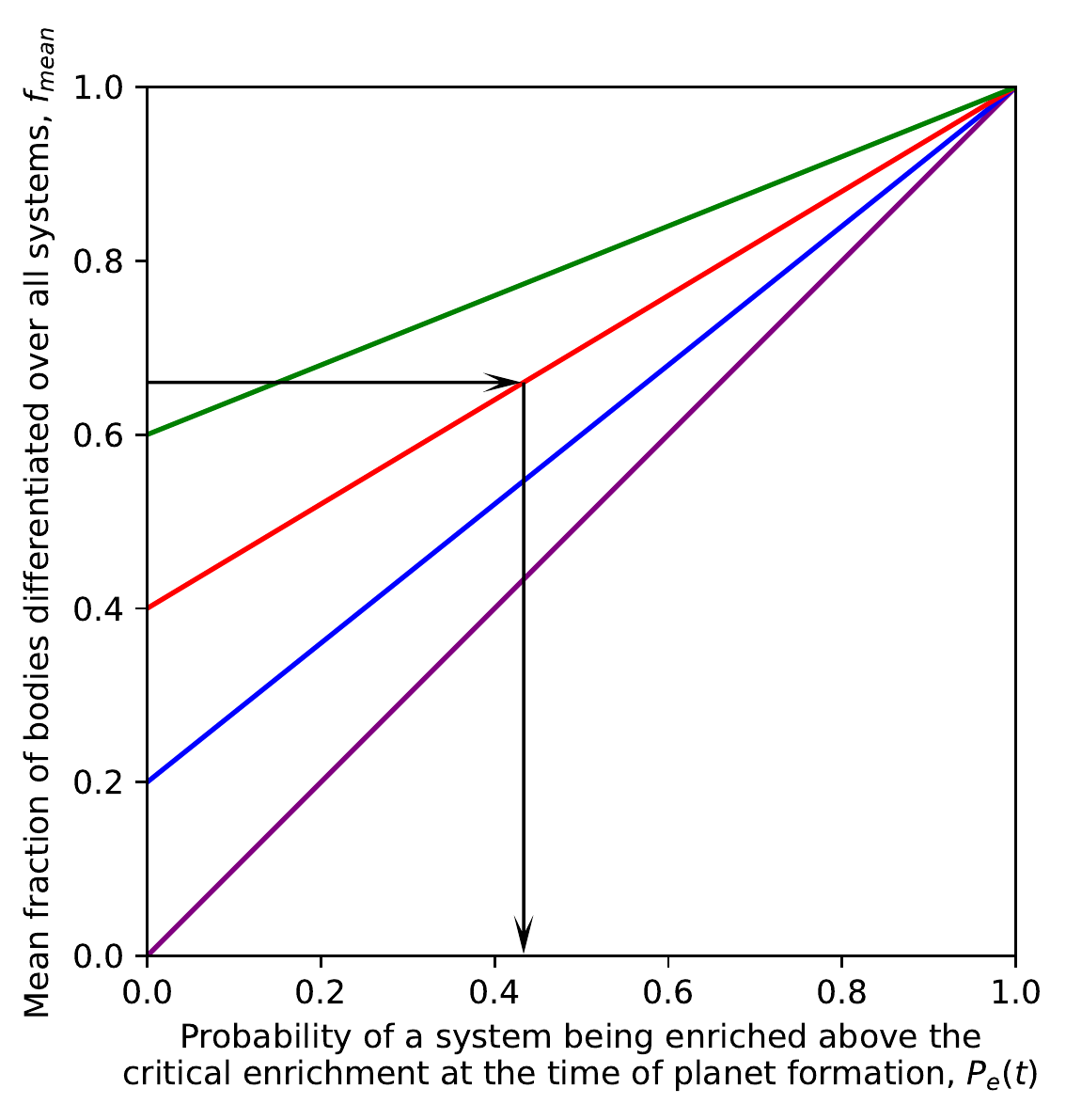}
	\caption{Plot of the mean differentiation fraction of systems $f\mean$ as a function of the probability that a system is enriched about the critical level, $P_e$, for different levels of differentiation with no enrichment, $f\unenr$(colours). $f\unenr$ for each colour is given by the intercept with the $y$-axis. The arrows demonstrate how this plot can be used to infer $P_e$ with knowledge of $f\mean$ (e.g. from WDs) and $f\unenr$ (e.g., from knowledge of the initial size distribution of planets). }\label{fig: fmeanPe}
\end{figure}

The degeneracy between heating due to gravitational potential energy and heating due to the decay of \SLRs\ is visualised in Fig. \ref{fig: fmeanPe}. The same mean fraction ($f\mean$) occurs for different combinations of $f\unenr$ and $P_e(t)$.

\subsection{Linking the fraction of bodies that are differentiated to their sizes}\label{sec: frac_to_sizes}
\begin{figure}
	\centering
	\includegraphics[width = \linewidth]{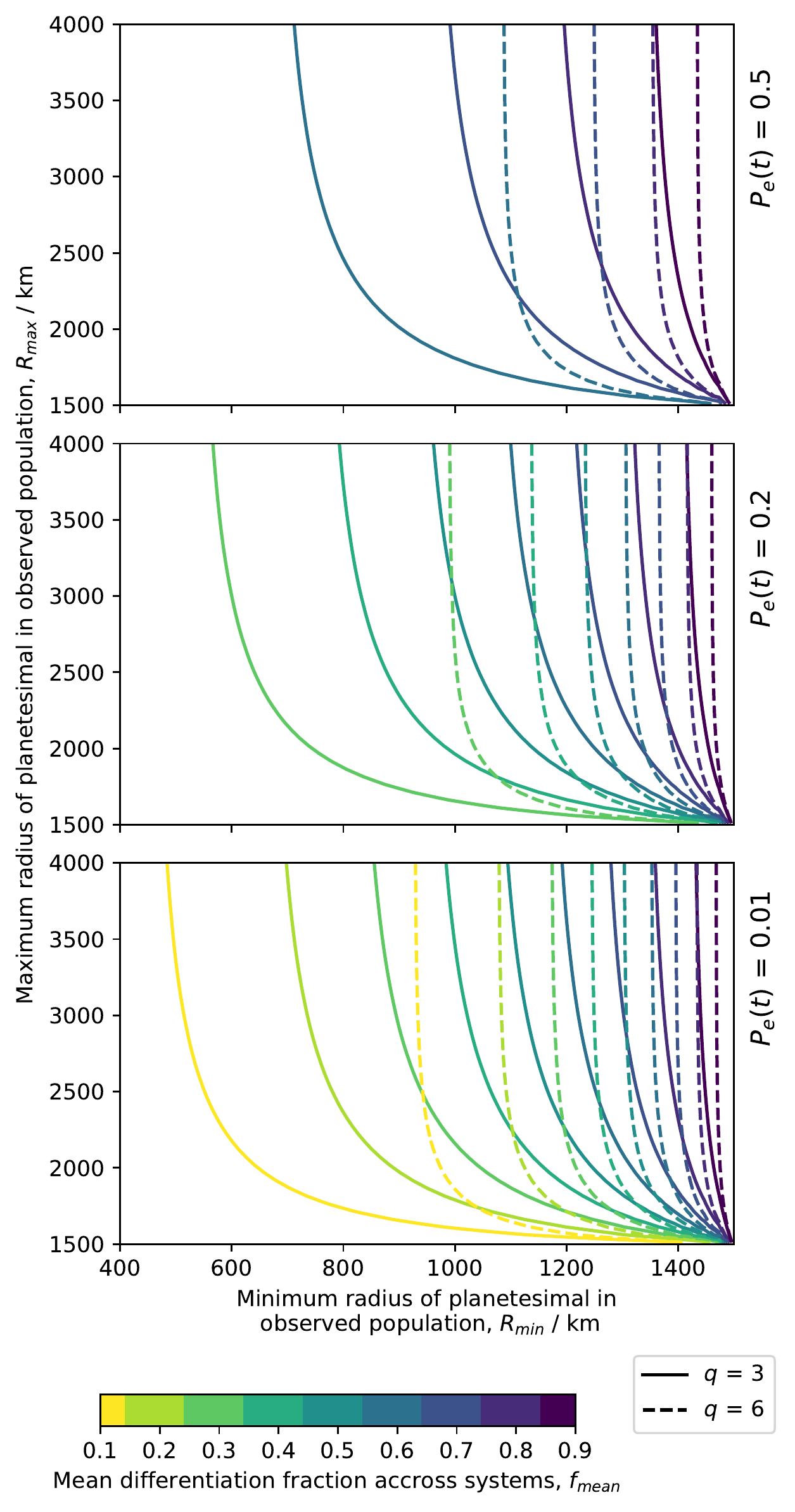}
	\caption{Contours of the mean proportion of bodies differentiated across all systems, $f\mean$, in terms of the minimum and maximum radii of planetesimals that contribute to the distribution. This is shown for different probabilities of a system being enriched above the critical level of enrichment at the time of formation, $P_e(t)$, and different slopes of the size distribution, $q$. The values of $q$ chosen are motivated in \S\ref{sec: size_dist}. }\label{fig: contours}
\end{figure}

Assuming the size distribution in \S\ref{sec: size_dist} then 
\begin{equation}
	f\unenr = \frac{\int^{R\Max}_{R\acc} r^{-q} \mathrm{d}r}{\int^{R\Max}_{R\Min} r^{-q} \mathrm{d}r} = \frac{1-(R\Max / R\acc)^{q-1}}{1-(R\Max / R\Min)^{q-1}},
\end{equation}
where both the numerator and denominator are negative since $q > 1$. Inserting this into Eq. \ref{eq: simplefmean} and rearranging yields
\begin{equation}
	R\Min =R\Max/\left(1 - \frac{1-P_e}{f\mean - P_e} \left(1-\left(\frac{R\Max}{R\acc}\right)^{q-1}\right)\right)^{\frac{1}{q-1}} \label{eq: Rmin_rmax}
\end{equation}
This, as expected, is undefined for $R\Max \leq R\acc$, i.e. where $f\unenr = 0$. It is also invalid for $f\mean \leq P_e$, as such a solution clearly cannot solve Eq. \ref{eq: simplefmean}. Plots of $f\mean$ as a function of $R\Max$ and $R\Min$ for some different values of $q$ and $P_e(t)$ are shown in Fig. \ref{fig: contours}. This demonstrates the intrinsic degeneracy of $R\Min$ and $R\Max$. The upper and lower bounds of the distribution (Fig. \ref{fig: cartoon}) can be squeezed or stretched while giving the same fraction. 

\begin{figure}
	\centering
	\includegraphics[width =\linewidth]{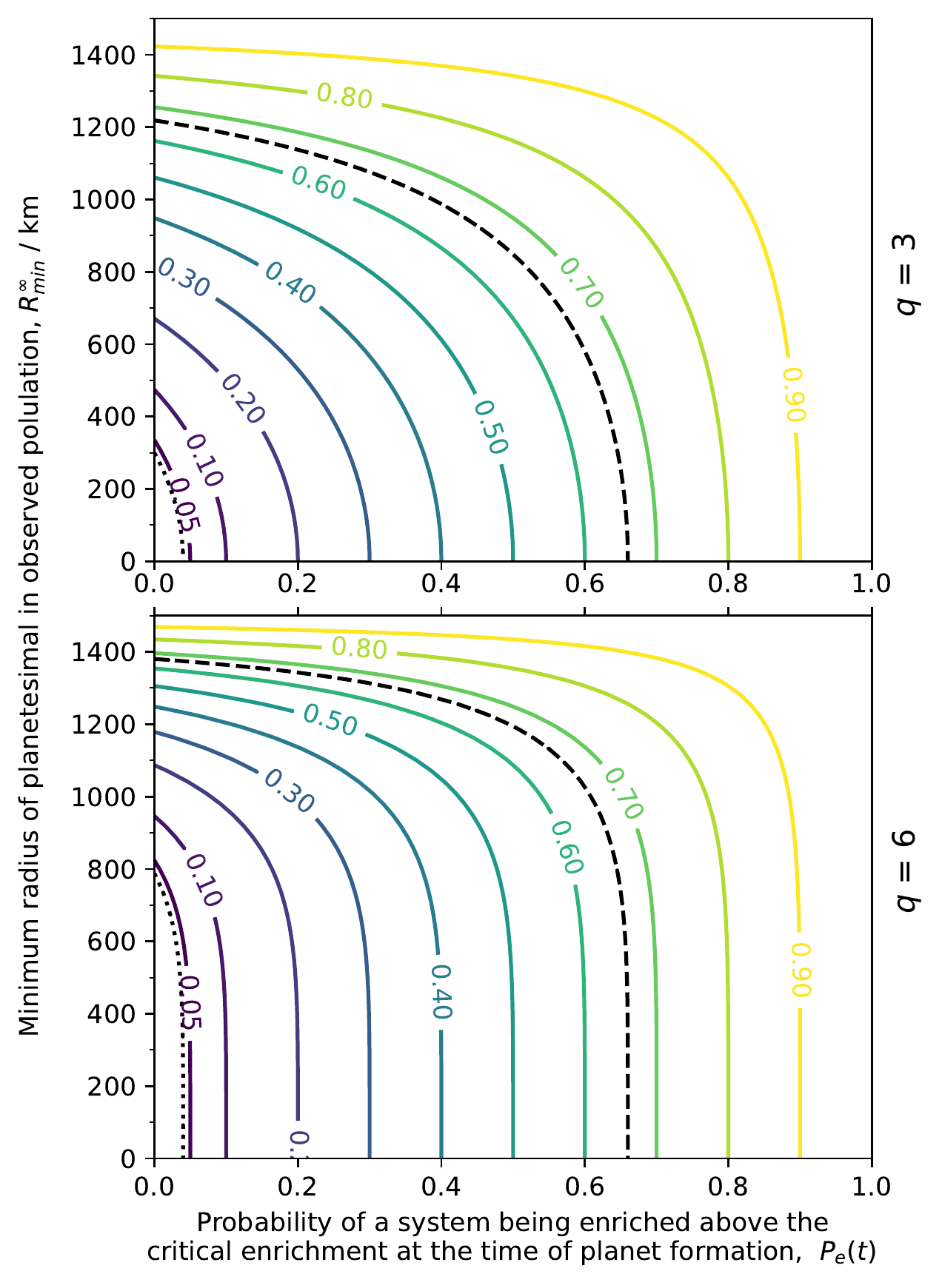}
	\caption{Contours of the mean differentiation fraction of systems, $f\mean$, as a function of the minimum size of bodies, $R\Min^\infty$, and the probability that a system has enrichment above a critical level, $P_e(t)$. Different panels show different slopes of the size distribution, $q$, the values of which are motivated in \S\ref{sec: size_dist}. The black dotted and dashed lines mark the curves for 4\% and 66\% referenced in \S\ref{sec: consequences} }\label{fig: rmins}
\end{figure}

Taking $R\Max \rightarrow \infty$ allows this interdependence to be removed:
\begin{equation}
	R\Min \rightarrow R\acc \left(\frac{f\mean - P_e}{1 - P_e}\right)^{\frac{1}{q-1}} \equiv R\Min^\infty \label{eq: Rmin_inft}
\end{equation}
This means that the dependence of $R\Min$ on $P_e$ can be seen more easily in a single figure (Fig. \ref{fig: rmins}). Furthermore, we would expect the largest planets to be much larger than the smallest. Importantly, it is also the lowest possible value $R\Min$ can be, which will be helpful in interpretation.

\subsection{Summary and use of our model}

{We have now explained the logic in our model that links the level of enrichment by \SLRs\: in forming planetary systems to the frequency of iron core formation in rocky bodies. The model assumes that there is a distribution of \SLRs\: across exoplanetary systems (Fig. \ref{fig: fxform}, Panel B). In planetary systems with enrichment above a critical level, $x_{\rm crit}$, which is all rocky planetary bodies considered here, and that form sufficiently early, an iron core will form (Fig. \ref{fig: fxform}, Panel A)}. In those planetary systems where rocky planetary bodies form later ($t_{\rm form} \gtrsim$ 1--2 Myr), a lower fraction of them will form an iron core. In all planetary systems, including those with no enrichment ($x<x_{\rm crit}$), the largest planetary bodies (Plutos) will form an iron core due to gravitational potential energy alone. The fraction of planetary bodies in a given system that are sufficiently large, $f\unenr$, is a function of the size distribution, and can either be considered in detail (as in \S\ref{sec: frac_to_sizes}) or used itself as a parameter (as in Eq.\ref{eq: simplefmean}, Fig. \ref{fig: fmeanPe} and Fig. \ref{fig: fxform}, Panel A), which can vary from 0 to some (small) fraction. {The product of the enrichment probability (B) and the distribution of formation times and the fraction of planetary bodies that are Plutos (A), gives the distribution over all systems (C), which is potentially bimodal.}

The white dwarf observations sample the mean fraction ($f\mean$) of planetary bodies forming an iron core across all exoplanetary systems. If the fraction of bodies that would be differentiated in non-enriched systems, $f\unenr$, i.e., Plutos, can be estimated, then Fig. \ref{fig: fmeanPe} (and equivalently Eq. \ref{eq: simplefmean}) can be used to estimate the probability of enrichment, as demonstrated by the black arrows. In the example shown in Fig. \ref{fig: fmeanPe} we have taken 66\% of systems to be recording accretion of differentiated material and the differentiation proportion of bodies in un-enriched systems to be 40\%; from this we can infer that $\sim 43\%$ systems were enriched with \SLRs\:  in their early history. Alternatively, $P_e(t)$ might instead be calculated through modelling or observations of star-forming regions \citep[e.g.,][]{2020RSOS....701271P,ReiterAl26} and used in combination with $f\mean$ to predict $f\unenr$. An identical principle can be applied to Eq. \ref{eq: Rmin_rmax} (Fig. \ref{fig: contours}) and Eq. \ref{eq: Rmin_inft} (Fig. \ref{fig: rmins}). 

\subsection{Are differentiated bodies due to Plutos or due to isotope enrichment?}\label{sec: pluto_or_SLR}

Rocky bodies with an iron core observed polluting white dwarfs may have differentiated due to having received a large amount of energy from accretion or through radioactive decay, as seen in Eq. \ref{eq: simplefmean} and Fig. \ref{fig: fmeanPe}. Without knowing how large the original body was, there is no way of distinguishing between these two scenarios, particularly in the case of polluted white dwarfs, where the white dwarf has likely accreted a small fragment of a parent planetary body that was originally significantly larger. 

The same mean fraction of rocky bodies that form an iron core ($f\mean$) can, therefore, occur in two ways. Firstly, a size distribution could be skewed to large planetary bodies (low $q$). Secondly, a planetary system could have a higher level of enrichment (high $x$). Eqs. \ref{eq: Rmin_rmax}, \ref{eq: Rmin_inft} and Fig. \ref{fig: rmins} quantify the size range of parent bodies that would have to be sampled, assuming a power law size distribution, in order to produce an observed $f\mean$, given a certain probability of \SLR enrichment, $P_e(t)$. If one is to not attribute most of the differentiation to \SLR\: heating, one has to be sampling large bodies with radii of 100s kms. One can see this in Fig. \ref{fig: rmins}: for a given mean differentiation fraction (coloured curves) it tends towards the probability of enrichment above the critical enrichment -- i.e., the curves are close to vertical in the bottom region of the diagram approaching the horizontal-axis.

\begin{figure}
    \centering
    \includegraphics[width =\linewidth]{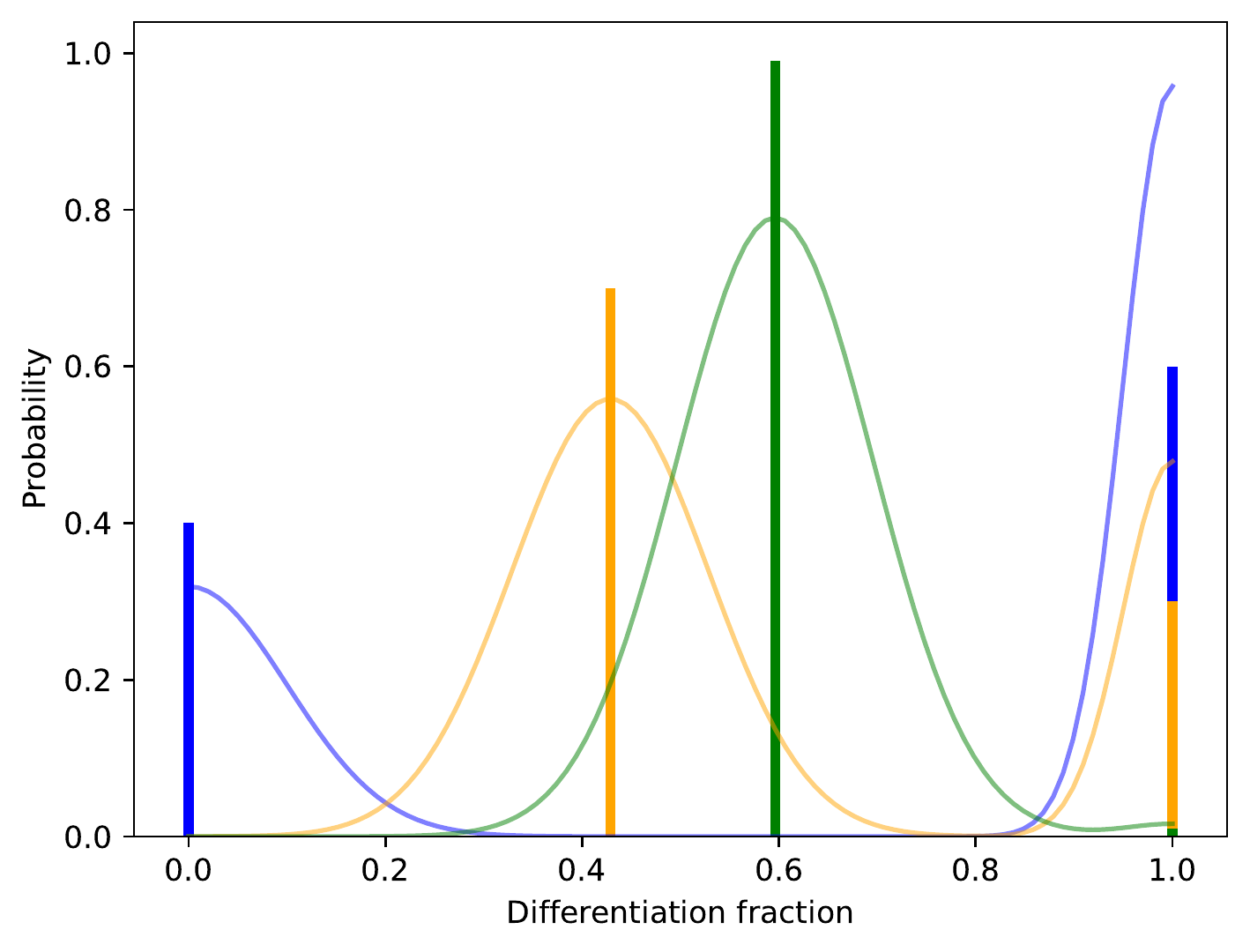}
    \caption{{Three different distributions of the fraction of systems that are differentiated over all systems that give the same mean of $f\mean = 60\%$. Bars show the simple picture described in our model, while the curves represent pictorially, what might happen when inter-system variation is included (\S\ref{sec: real}). In green, large bodies are common, and enrichment is uncommon in the white dwarf pollutant population, meaning all systems have the same differentiation fraction. In blue is the other extreme, where for un-enriched systems no bodies are differentiated, and $f\mean$ is entirely generated through fully differentiated systems. Orange shows an intermediate case.}}
    \label{fig: samefmean}
\end{figure}

Another demonstration of this can be seen in Fig. \ref{fig: samefmean}. The same value of $f\mean$, can be generated by different combinations of systems with high or low numbers of differentiated bodies.
\section{Linking white dwarf observations to exoplanet enrichment levels}
\label{sec: linking_evidence}

\subsection{Polluted white dwarf constraints on core-mantle differentiation} \label{sec: how_much_diff}

The best observational signature of core--mantle differentiation in exo-asteroids comes from the composition of planetary material accreted by white dwarfs. Whilst most white dwarfs have accreted primitive, rocky material, whose abundances lie within a small range around solar \citep{JuraYoung, Harrison18}, the material in the atmospheres of some white dwarfs exhibits extreme differences between the abundances of siderophile (Fe, Ni, Cr) and lithophile (Ca, Mg, Si) species. Some compositions have been found to be inconsistent with primitive rock, and are better explained as fragments of planetary cores or mantles. 

Ideally, to determine how often exo-asteroids form an iron core, an unbiased sample is required. Even within the Solar System (i.e., the meteorite collection) such an unbiased sample does not exist \citep{2020SSRv..216...27M,2021SciA....7J7601B}. For exoplanetary systems the task is even harder. Polluted white dwarfs sample a selection of asteroids that were scattered inwards from outer planetary systems. Whilst the scattering process is unlikely to be significantly biased towards the core--mantle status of a planetesimal, there are significant observational biases in any sample of polluted white dwarfs. The 202 white dwarfs from \cite{Hollands2017, Hollands2018} analysed in \cite{Harrison2021-Bayes} provide a sample that were selected in a uniform manner and have SDSS spectra with similar (although not the same) S/N. However, the colour cuts used to select these white dwarfs potentially introduce biases towards particular abundances \cite[see][]{Harrison2021-Bayes}. \citet{Harrison2021-Bayes} find 8/202 (4\%)  to have accreted core-rich fragments to $>3\sigma$. 

It is easier to probe core--mantle differentiation when several sidereophile and several lithophile species are detected, compared to the 3 elements detected for the majority of white dwarfs in \cite{Hollands2017, Hollands2018}. However, whilst a sample of 42 white dwarfs with >5 elements can be collated \citep[e.g.,][]{2022MNRAS.510.3512B}, these white dwarfs were not selected in any uniform manner, nor observed in a similar manner. Nonetheless \cite{2022MNRAS.510.3512B} find that 16/42 (40\%) of these white dwarfs show evidence for core or mantle-rich compositions, suggesting that the fraction of core--mantle differentiated white dwarf pollutants could be significantly higher, with many core-rich fragments not identified from the abundances of only Ca, Mg and Fe in the \cite{Hollands2017, Hollands2018} sample. 

It is important to note that differentiation is only evident in polluted white dwarfs when they have accreted a core- or mantle-rich fragments: a differentiated body that is wholly accreted to the white dwarf, with the exception of possible volatile element loss, will have the same composition as a primitive undifferentiated object.  As a result, the observed core- and mantle-rich fragments may only be the extreme tails of a larger distribution, such that the real fraction of core--mantle differentiated white dwarf pollutants is significantly above the observed fraction. This would be the case if white dwarf pollutants sample the results of many collisions between core--mantle differentiated planetesimals \citep{Marcus2009,2015Icar..247..291B}. This distribution would then contain many planetesimals with core mass fractions close to initial values, such that the composition of the entire fragment is the same as that of a primitive body \citep{Bonsor2020}. \cite{Bonsor2020} use the 203 white dwarfs from the \cite{Hollands2017, Hollands2018} sample to estimate that more than two thirds, if not all, planetesimals accreted from white dwarfs are the fragments of parent bodies that formed an iron core. 

To summarise, the observations cannot yet provide a definitive answer as to how commonly exoplanetesimals are differentiated. However, fraction of core--mantle differentiated white dwarf pollutants is at least 4\%, most likely tens of percent and could even be as high as two thirds. 

\subsection{Polluted white dwarf constraints on enrichment with short-lived radionuclides}

\label{sec: consequences}

For any inference about the level of core--mantle differentiation across exoplanetary systems, there is a degeneracy between white dwarf pollution occurring from the accretion of large planetary bodies (>Pluto) or high levels of \SLR\: enrichment. The conservative estimate (see \S\ref{sec: how_much_diff}) of at least 4\% of exoplanetary bodies accreted by white dwarfs being the fragment of core--mantle differentiated planetesimals, can in principle be explained by very low levels of \SLR\: enrichment across exoplanetary systems. However, this applies only if the population of planetary bodies accreted by white dwarfs samples a size distribution with minimum radius of $\gtrsim \SI{300}{\km}$ (Fig. \ref{fig: rmins}, dotted line) and those individual white dwarfs with core or mantle-rich compositions accreted fragments of Pluto or larger-sized bodies. This would mean that all white dwarfs accreted bodies on the order of the largest asteroids, (e.g., Vesta, Pallas).

Alternatively, if the higher estimate that at least two thirds of planetary bodies in WDs are differentiated is closer to the true underlying distribution of the observations \citep{Bonsor20}, one deduces from Fig. \ref{fig: rmins} (black dashed line) that unless the proportion of enriched systems is above $\sim 50\%$ then the smallest bodies in WDs must be $\gtrsim \SI{1000}{\km}$. Considering that the largest body in the asteroid belt, Ceres, is only $\sim\SI{470}{\km}$ \citep{Ceres2016}, this seems implausibly large as a minimum size of white dwarf pollutant. Therefore, a high estimate of the fraction of differentiated bodies supports \SLR\: enrichment scenarios in which a significant fraction of planetary systems are enriched, on the order of tens of percent.

It is also worth noting that our estimate for the radius at which planetesimals can differentiate due to gravitational energy is likely a lower limit (\S\ref{sec: R_acc}). Were it higher, the minimum radius of white dwarf pollutant progenitors would also have to increase, making the sizes required even more implausible.

\subsection{How commonly are exoplanetary systems enriched by short-lived radionuclides?} \label{sec: Al-orgin}
{For this section, we focus on the knowledge of the prevalence of \isotope[26]{Al} since it made the most significant contribution to radiogenic heating in the early solar system, and has thus been the most studied. We will return to a broader picture of \SLR\: heating at the end of the section.}

The enrichment of the Solar System  with \isotope[26]{Al} has often been attributed to supernovae or AGB stars, but more recently mass loss from the winds of massive (generally Wolf-Rayet) stars have gained interest. Wolf-Rayet (WR) stars offer a promising origin for the Solar System specifically, because supernovae produce an overabundance of the \SLR\: \isotope[60]{Fe} relative to the Solar System \citep{Gaidos09,GounelleM12}, although uncertainties in production and incorporation rates are large enough for supernovae not to be excluded \citep[][]{2014ApJ...789...86A,Licht16-SN}. Incorporation into the planetary system can be either through injection into the molecular cloud that went on to form the Solar System (e.g., \citealt{Cameron77,2016ApJ...826...22K,2019ApJ...870....3B} for supernovae; \citealt{Gaidos09,GounelleM12,2017ApJ...851..147D} for WR stars) or later to the protoplanetary disc (e.g., \citealt{Chevalier2000,Licht16-SN} for supernovae; \citealt{Portegies19} for both).

Each mechanism has a different probability of Solar System enrichment levels. \citet{Gounelle15} suggest $\sim 1\%$ of systems should have enrichment of \isotope[26]{Al} above Solar System levels, by considering the chances of a finding cluster with appropriate properties \citep[though see][about the triggering hypothesis]{2016MNRAS.456.1066P}, while \citet{Portegies19} estimate there to be only 36,000 Solar System-like systems in the Milky Way, through arguments about the rate that systems will form under their proposed scenario. 

Alternatively, \isotope[26]{Al} could, in fact, be common in star forming regions. Local \citep[][]{vasil13-GMC,2016ApJ...826...22K,2014E&PSL.392...16Y,Young16-Bayes} and galactic-scale  \citep{2020MNRAS.497.2442F} hydrodynamical models of star-forming regions suggest sufficient mixing of isotopes produced by SNe and WR stars alongside new star/planetary system formation, even though so far none of these models can reproduce the relative fractions of all relevant isotope systems \citep{lugaro18-review,2021PASA...38...62D}. Observational evidence from gamma rays produced in \isotope[26]{Al} decay suggests that star forming regions retain \isotope[26]{Al} beyond its half-life meaning it is replenished by multiple sources \citep{ReiterAl26,2021NatAs...5.1009F}. In this case, the probability of enrichment is linked to whether a star formed in a sufficiently dense cluster. \citet{ReiterAl26} estimate that $\sim 25\%$ of systems are enriched in \isotope[26]{Al} to a similar level as the Solar System, based on this argument.

{We have focused on \isotope[26]{Al}, however any \SLR\: must be produced by a relatively recent stellar event, either a supernova or in a massive star, thus the arguments of there needing to be chance encounters or mixing in star forming regions to supply \SLRs apply more broadly. A possible adjustment to the modelling would be if supernovae were to contribute significantly to a system's \SLR\: inventory: in this case \isotope[60]{Fe}, will become an important source of heat due to its lack of production in WR winds \citep{Woosley2007,Gaidos09}, although it is unlikely to entirely change the picture. }

\subsection{Summary of observational constraints}

The white dwarf observations point towards a minimum of 4\% of exoplanetary systems having exoplanetesimals that differentiated to form an iron core and plausibly a significantly higher fraction, on the order of tens of percent. Whilst these observations could in principle be explained by the accretion of large (Pluto or bigger) planetary bodies, where gravitational potential energy alone fuelled the large-scale melting, the models presented here show that this would require all white dwarfs to be polluted by large asteroids, as well as Plutos to exist in many exoplanetesimal belts. However, the observational evidence is against the existence of Plutos in most exoplanetesimal belts, including our own asteroid belts (see discussion in \S\ref{sec: could_large}), meaning it is unlikely that a large proportion of the core--mantle differentiation can be attributed to large bodies. As the white dwarf observations point towards tens of percent of exoplanetesimals forming iron cores, it is more likely that a significant fraction (tens of percent) of exoplanetary systems were enriched in \SLRs. This in turn, points towards a ubiquitous pathway to \SLR\: enrichment across the galaxy. Whilst there is insufficient evidence to categorically rule out chance-encounter mechanisms for enrichment, such as single supernovae in low- to mid-sized star-forming regions, this work suggests that these mechanisms are not sufficiently efficient at enriching exoplanetary systems to explain the full population, as sampled by white dwarfs.

\section{Discussion} \label{sec: discuss}

This work presents an analytic framework that links observations of core or mantle-rich material in the atmospheres of white dwarfs to the level of enrichment by \SLRs\:across exoplanetary systems. This model relies on the interpretation of state-of-the-art observations to estimate the fraction of bodies accreted by white dwarfs that are fragments of core--mantle differentiated planetary bodies. Such inferences will become more reliable in the future. Current estimates point towards a significant fraction (tens of percent) of exoplanetary systems being enriched in \SLRs, which in turn points towards ubiquitous mechanisms for \SLR\:enrichment, rather than chance encounters with supernova, WR stars, or isolated AGB stars. The most significant caveats to this conclusion are the reliability of the white dwarf observations and whether all core or mantle-rich white dwarf pollutants could indeed be fragments of planetary bodies sufficiently large that gravitational potential energy alone could lead to enough heating to form an iron core. 

\subsection{Could the white dwarf pollutants be large?}\label{sec: could_large}

A key degeneracy exists, whereby, if those white dwarf pollutants that exhibit signatures of core- or mantle-rich material are in fact large planetary bodies, or indeed fragments of large planetary bodies, there is no need to invoke \SLRs\:as gravitational potential energy alone can fuel the large-scale melting and core formation (see detailed discussion in \S\ref{sec: pluto_or_SLR}). Whilst this is plausible, particularly if the fraction of white dwarfs with core or mantle-rich compositions is low, it is important to ask the question of whether such large bodies ($R\gtrsim\SI{300}{\km}$) exist often enough in planetesimal belts. Whilst the Kuiper belt clearly contains a number of Pluto-sized objects, such bodies are rare in the asteroid belt and there is evidence from the decay of infrared emission with time in debris discs around A stars \citep{wyattdebris}, and the mass budgets, compared to planet-forming discs  \citep{Krivov2021} that suggests that such large bodies do not exist in most planetesimal belts. {Furthermore, to produce core-rich fragments of these objects they must have undergone a major collision with a similarly high mass object, which makes them being the origin of the differentiation signatures more unlikely.} Thus, whilst we cannot rule out that a small fraction of white dwarfs are indeed polluted by planetary bodies sufficiently large to form an iron core without \SLR\:enrichment, this is unlikely to explain a large (tens of percent) fraction of white dwarfs with core or mantle-rich compositions. 

{On the other hand it it is possible that even if such disruptions are rare they could generate a large amount of material, allowing higher proportions of core-rich fragments to be observed. To investigate this further detailed consideration of the dynamics and collision physics would need to be considered.} 

In principle, estimating the size of a white dwarf pollutant is possible on the basis of its geochemistry alone.  \citet{2022MNRAS.510.3512B} showed that the pressure-dependence of element partitioning between core and mantle during differentiation can provide a compositional fingerprint of the size of a body.  To apply this to white dwarfs requires high-precision measurements of elements such as Cr and Ni, which exist only for a few systems at present.  In future though, the method described in \citet{2022MNRAS.510.3512B} offers a route to ruling in or out the accretion of planet-sized objects on a system-by-system basis.

\subsection{Reliability of the white dwarf observations and prospects for the future} \label{sec: WD_reliable}
 A key limitation of this work is the existence of sufficiently large, unbiased, sets of polluted white dwarfs, where a sufficient number of elements are detected in sufficiently high S/N spectra that the abundances can be accurately constrained.
 
 Another important limitation is the interpretation of the observed abundances in white dwarfs. Firstly, it is possible to alter abundances through differential sinking times, so careful modelling of how the abundances evolve is required \citep[e.g.,][]{Harrison18}. Additionally, mantle-rich material is difficult to distinguish from material that experienced heating and the loss of moderately volatile species, as both have elevated Ca/Fe ratios  \citep[see \S3.3 of][]{Harrison2021-Bayes}. 
 
Furthermore, often the quality of the data is insufficient to be certain that the pollutant is non-primitive. For instance, \cite{Harrison2021-Bayes} find that 68/202 WD abundances in their sample \citep{Hollands2017,Hollands2018} are best explained by core-rich fragments, yet for only 8 is this conclusion statistically significant ($>3\sigma$), reducing their estimate to 4\%. This is mostly due to degeneracies in interpreting abundances derived from low S/N data, where for most white dwarfs only 3 elements were detected in the photosphere. Significantly better inferences can be made for white dwarfs where many elements are detected, but biases, some of which are unresolved, exist in the full sample currently available.

{While \cite{Hollands2017,Hollands2018} is unbiased in its selection criteria, it is a sample of only He dominated DZ white dwarfs. The large convection zones of these white dwarfs result in relatively long sinking times $\sim \SI{1e5}{} - \SI{1e6}{yrs}$ \citep{Hollands2018} meaning the material in the white dwarf atmospheres can sample multiple accretion events, which results in the largest recently accreted body dominating. This means the sample may preferentially include larger objects. Since it is the size of objects in the initial population that is required to explain significant differentiation, not the size of the actual objects accreted to the white dwarf atmospheres, it is not immediately obvious how this bias affects our conclusions. On the one hand large pollutants might have to have been fragments of large progenitors, thus we would expect to be preferentially sampling large differentiated planets, not smaller planets that require radioactive decay to differentiate. This argument depends on how the fragmentation of the planets occurs. On the other hand, preferentially sampling larger bodies may make it even more unlikely to detect core-rich fragments, as they are necessarily smaller, increasing the inferred initial amount of differentiation further. Were this the case it may be more necessary to appeal to \SLRs, when following through the arguments in the rest of this work. More careful analysis, considering the balance between rarity of large bodies and their longevity in the atmospheres, is required to fully resolve this bias.}

Fortunately, the future prospects for improving our knowledge of the fraction of white dwarf pollutants that are core or mantle-rich are good, with large surveys such as WEAVE, 4-MOST, DESI and SDSS-V providing spectra of Gaia detected white dwarfs. Not only will the number of systems observed increase, but the biases in the samples will be better understood, such that, with care in interpreting the observations, the true population can be more tightly constrained. 

\subsection{Implications for exoplanets}

As outlined in the introduction, heating by short-lived radionuclides has profound implications for the geophysical and geochemical evolution of planetary objects during accretion. Specifically, its impact on the bulk volatile mass budget has the potential to rewrite our current understanding of volatile inheritance, which is largely based on distance to the central star. If present in a large fraction of exoplanetary systems, as suggested by our inferences in this work, this introduces radionuclide-driven internal heating as a first-order factor that distinguishes Solar System-like planetary systems from others, which would be reflected in the galactic exoplanet population. Systems heated by short-lived radionuclides would tend to form drier, and hence smaller exoplanets than their not-enriched counterparts \citep{2019NatAs...3..307L}. For not-enriched systems this bears the potential for very high water mass fractions up to several wt\% of their bulk mass \citep{2003ApJ...596L.105K,2004Icar..169..499L}. The concentration of water in particular has the effect of oxidizing planetary materials, which can decrease the propensity of fully-fledged planets to form iron cores \citep{2008ApJ...688..628E}. 

At present day, mature exoplanetary systems are probed only for short-period orbits and prevalently larger planets. Among super-Earths, the oxidation state of their surfaces and potential secondary atmospheres \citep{2021ApJ...909L..22K,2021ApJ...914L...4L,liggins2022} and the melting state of planetary interiors \citep{2021ApJ...922L...4D} may yield clues on the prevalence of volatile delivery and desiccation mechanisms among planetary systems, which are ultimately linked to the distribution of short-lived radionuclides across these systems. Additionally, observations of sub-Neptunes may be able to distinguish between water-enriched and water-poor planetary interiors \citep{2021ApJ...921L...8H,2021ApJ...922L..27T}. In the upcoming years, detailed inferences from select exoplanets (JWST, ELTs) and population statistics from transit surveys (e.g., TESS, PLATO) may thus be used to build a bridge from our understanding of planetary accretion to mature exoplanet systems using the inferences we have presented here in this work. A high fraction of short-lived radionuclide-enriched planetary systems predicts that the internal heating during planetary accretion would be imprinted into the planetary bulk volatile distribution of small exoplanets, with a moderate proportion ($\sim$10s\%) of systems being enriched and hence dry, and the others being not significantly enriched and hence more volatile-rich. Ultimately, this suggests a qualitative dichotomy between \SLR\: enriched and potentially habitable, and drowned, not-enriched exoplanetary systems. Large-scale space-based direct imaging surveys will be necessary to probe these distributions for terrestrial-like exoplanets on orbits beyond the runaway greenhouse threshold \citep{2019arXiv191206219T,2020arXiv200106683G,Quanz21}.

\subsection{Limitations} \label{sec: real}

\subsubsection{Timing of planetesimal formation}

The timing of planetesimal formation is crucial, as \SLRs, decay substantially over typical planet formation timescales. $^{26}$Al decays exponentially with a half-life of 0.72 Myr, such that planetesimals in systems with a Solar System-like initial enrichment level that form after ~1 Myr are no longer sufficiently enriched to form an iron core. In this simple model, all planetesimals are assumed to form at a single epoch. Under this simplification we predict either one or two distinct populations. Either no systems are enriched (or all planetesimals form late) meaning all systems have the same level of differentiation (green bar in Fig. \ref{fig: samefmean}). Alternatively, only some systems are differentiated, meaning there is a population of systems where all bodies are differentiated, and a population where fewer bodies are differentiated (orange and blue bars in Fig. \ref{fig: samefmean}). 

Planetesimal formation clearly continued over a significant time period in the Solar System; meteorite ages vary from CAIs to several Myrs after CAIs \citep{2020SSRv..216...55K,2020SSRv..216...27M} when considering both the inner and outer Solar System. {This results in populations with different properties even within the same system \citep[][]{2021Sci...371..365L,2022arXiv220310023L}.} The assumption of a single epoch, however, bears out when only the mean fraction of planetary bodies that form an iron core, $f_{\rm mean}$, across all exoplanetary systems is considered. If a distribution of formation times were to be considered, the strict delta function populations will be spread out (curves in Fig. \ref{fig: samefmean}), and the fraction of bodies that are differentiated in enriched systems will be below 1 because some bodies formed early and some formed late. 

From an observational perspective, early planetesimal formation is favoured by inferences of the depletion of mm-sized dust grains in ALMA disks \citep{2017AJ....153..240A,2020A&A...640A..19T}. If planetesimals form preferentially during the first few hundred thousand years of planetary system formation, the time delay between enrichment with short-lived radionuclides and planetesimal formation is of minor importance.

If there is any significant late formation, for instance if white dwarf pollutants preferentially sample wide orbit objects, however, it only strengthens our conclusions about the ubiquity of \SLR\:enrichment: the later objects form the less they are influenced by radioactive decay, so enrichment must be even more common, or objects even larger, to explain observed differentiation.

\subsubsection{The appropriate size distribution} \label{sec: approp_dist}

The conclusions presented here rely upon the size distribution of planetesimals in exoplanetary systems. This size distribution reflects both the distribution at formation and how material is drawn from it to be scattered onto the white dwarf. Thus, by using a simple power law, we are simplifying the further dynamic evolution of the system. Detailed modelling of this is beyond the scope of this work, and indeed the size distribution likely varies between exoplanetary systems, resulting in a system to system spread in the fraction of bodies with cores (Fig. \ref{fig: samefmean}).

Here, we aim instead to capture the broader picture by considering ranges in the parameters of the power law size distribution. The slope of the power law, $q$, is less critical than the minimum and maximum planetesimal radii, as can be seen in Figs. \ref{fig: contours} and \ref{fig: rmins}. Since the model is kept relatively simple, all we can infer is that regardless of $q$ in order to explain high mean differentiation fractions (e.g., $\sim 60\%$) with low enrichment probabilities (e.g., $\sim 1\%$) one needs large bodies ($\gtrsim \SI{300}{\km}$). 

The values of $q$  that we use are in agreement with current predictions of the streaming instability \citep{2019ApJ...885...69L} and birth size frequency distributions derived from the asteroid belt \citep{2017Sci...357.1026D}.

\subsubsection{Validity of a critical enrichment value}\label{sec: valid_crit}

In our model we use a critical value of the \isotope[26]{Al} enrichment level above which bodies differentiate. As explained in \S\ref{sec: R_SLR} this is justified if we only consider bodies $\gtrsim \SI{50}{\km}$. It is difficult to find {\it a priori} estimates for the minimum sizes of white dwarf pollutants, because it requires modelling of the specific atmosphere and in most cases the body is still accreting, meaning an estimate of the mass in the atmosphere is only a minimum (although see discussion in \S\ref{sec: could_large}). Best estimates, using non-accreting scenarios \citep{Harrison2021-Bayes}, are that the WD pollutants are $\gtrsim \SI{80}{\km}$. Furthermore, the bodies in WDs that show extreme abundances are potentially fragments of larger bodies (see \S\ref{sec: how_much_diff}). Thus it is likely that our assumption is justified. However, if the white dwarf sample does have smaller minimum radii, it simply supports our conclusion that \SLR\: enrichment is common, because even higher enrichment would be required to explain the core--mantle differentiation if objects lower minimum radii ($R\Min$) are present.

\section{Conclusions} \label{sec: conclusion}

In the Solar System, the presence of \SLRs, notably $^{26}$Al, fuelled the large-scale melting and formation of an iron core in rocky planetesimals as small as $\sim$10 km. White dwarfs that have accreted planetary material provide an opportunity to study the formation of iron cores in exoplanetesimals and investigate whether exoplanetary systems are commonly enriched in \SLRs. Here we have shown that, if a large fraction of white dwarfs have accreted the fragments of planetary bodies that differentiated to form an iron core, these observations indicate that either most white dwarfs accrete fragments of bodies Pluto-sized or larger, or that a large fraction of exoplanetary systems are enriched with \SLRs. We argue that the latter is more likely, as the low accretion rates and the rare dynamical pathways for moons or minor planets render the chance of all white dwarf pollutants being sufficiently large low.

This would suggest that the Solar System is not unusual in being enriched in \isotope[26]{Al} and points towards a ubiquitous pathway leading to the enrichment of exoplanetary systems, rather than a rare chance encounter with a single nearby supernova, Wolf-Rayet or AGB star. More detailed observations of differentiated material accreted onto polluted white dwarfs, modelling and observations of star-forming regions, and observations of the bulk volatile distribution and oxidation state of short-period, rocky exoplanets will reveal further insights into this key question that separates Solar System analogues from other exoplanetary systems.

\section*{Acknowledgements}
Insightful and important discussions with Richard Parker, Marc Brouwers and Andrew Buchan are acknowledged here. {We thank the reviewer for their helpful comments and suggestions.} A.C. acknowledges support of an STFC PhD studentship. AB acknowledges the support of a Royal Society Dorothy Hodgkin Research Fellowship, DH150130 and a Royal Society University Research Fellowship, URF\textbackslash R1\textbackslash 211421.
 T.L. was supported by a grant from the Simons Foundation (SCOL Award No. 611576).

\section*{Data Availability}
Data from the 2-D planetesimal evolution models presented in Fig. 1 are available in a public GitHub repository (\url{https://github.com/timlichtenberg/plts_evolution}). All other figures are reproducible from the equations provided in the text.



\bibliographystyle{mnras}
\bibliography{refs} 


\bsp	
\label{lastpage}
\end{document}